\documentclass[amsmath,amssymb,prl,twocolumn,superscriptaddress]{revtex4}
\usepackage{graphicx}
\usepackage{dcolumn}
\usepackage{bm}

\begin{document}

\title{Symmetry-Breaking Motility}

\author{Allen Lee}
\affiliation{Department of Physics, Massachusetts Institute of Technology,
Cambridge, Massachusetts 02139}
\author{Ha Youn Lee}
\affiliation{Department of Physics, Massachusetts Institute of Technology,
Cambridge, Massachusetts 02139}
\affiliation{Department of Physics, The Ohio State University, Columbus, Ohio
43210}
\author{Mehran Kardar}
\affiliation{Department of Physics, Massachusetts Institute of Technology,
Cambridge, Massachusetts 02139}

\date{July 29, 2004}

\begin{abstract} 

Locomotion of bacteria by actin polymerization, and \textit{in vitro}
motion of spherical beads coated with a protein catalyzing
polymerization, are examples of active motility. Starting from a
simple model of forces locally normal to the surface of a bead, we
construct a phenomenological equation for its motion. The singularities
at a continuous transition between moving and stationary beads are
shown to be related to the symmetries of its shape. Universal features of
the phase  behavior are calculated analytically and confirmed by simulations. 
Fluctuations in velocity are shown to be
generically non-Maxwellian and correlated to the shape of the bead.

\end{abstract}


\maketitle

Motility, the capacity of organisms for independent motion, has long
fascinated physicists as well as biologists. 
For example, the reversible character of low Reynolds number flows~\cite{purcell}
requires ingenious methods for swimming motion~\cite{golestanian}.
Another strategy for movement, the subject of
this paper, is propulsion via forces generated by polymerization of
actin on an organism's surface. 
Polymerizing actin filaments are involved in, among other things, the
locomotion of certain bacteria, such as the pathogenic \textit{Listeria monocytogenes}. 
Upon entering the cytoplasm of a host, a \textit{Listeria}
``hijacks" the cell's actin machinery; first a roughly symmetrical
cloud of polymerizing actin filaments envelops the bacterium, and
then an asymmetric ``comet tail" of actin spontaneously develops that
propels the bacterium forward~\cite{portnoy:jcb89}.  Motility by actin
polymerization has been the subject of several analytical and model
studies~\cite{oster:biopj93,oster:biopj96,prost:biopj00,carlsson:biopj01,oster:biopj03}.

A number of \textit{in vitro} experiments have been performed on
artificial systems designed to mimic \textit{Listeria} motility under
simplified and more easily controlled
circumstances~\cite{welch:currbio99,sykes:nat02,
pantaloni:jcb03,carlier:jcb03,theriot:pnas99}.  In these experiments,
the bacteria are replaced with spherical beads coated with a protein
that catalyzes actin polymerization. When these functionalized beads
are placed in cell extracts or a mixture of purified proteins,
phenomenology similar to that of \textit{Listeria} motility is
observed. A spherically symmetric cloud of cross-linked actin
filaments forms around each bead. Then, depending on the various
parameters of the experiment, some beads are observed to spontaneously
break free of their actin clouds and move away, propelled by an
asymmetric actin comet tail.

Several theories have been put forward to explain the mechanism behind
this spontaneous motility. van Oudenaarden and Theriot model the actin
filaments surrounding the bead as elastic Brownian
ratchets~\cite{theriot:ncb99}. The stochastic growth of these filaments
is coupled due to the presence of the bead; when one filament grows,
it gives the bead a small push, making room for neighboring filaments
to grow. This cooperativity generates positive feedback and the
possibility of an avalanching symmetry-breaking event. In another
model, Noireaux et al have proposed that the actin cloud surrounding
the bead be treated as an elastic
gel~\cite{sykes:biopj00}. Polymerization occurs at the bead surface,
which expands the gel and stretches it. At a certain point, stress
at the outer surface ruptures the gel, and the bead is free to
escape. Mogilner and Oster have suggested a third alternative in which
movement occurs because of the failure of cross-links within the actin
network~\cite{oster:biopj03}.

Our aim is not to reconcile these existing theories, or to provide a
comprehensive description of the observed phenomena. Instead, we
investigate the behavior of an {\em activated rigid bead} within a
framework that is sufficiently general to include many possible
microscopic models.  In the spirit of the Landau theory of magnetism, we
treat the actin filaments surrounding the bead as a continuous
effective field that generates a normal force density at each point on
the bead surface. The evolution of this order parameter field is
coupled to the velocity of the bead, allowing the possibility of
positive feedback as in the  above Brownian ratchet
model~\cite{theriot:ncb99}. Our approach neglects many aspects of the
system, but can provide answers regarding generic features: What are
the possible singularities that can accompany the transition between
stationary and moving states, and how do they depend on the shape of
the bead?  Regarding the active bead as a stochastic system, are there
characteristics of its fluctuations that set it apart from a passive
(Brownian) bead?

We consider a bead of fixed shape, with unit normal vector
$\hat{\mathbf{n}}(\mathbf{r})$ at a point $\mathbf{r}$ on its surface.  The
polymerization of actin (or any other actively energy-consuming
process for that matter) is assumed to exert a force, locally directed
along the normal, at each point on the bead surface. The force per
unit area at time $t$ is indicated by a scalar field $g(\mathbf{r},t)$,
which could for example be a function of the actin density at that
point.  The net force due to such activity, acting on the center of
mass of the bead, is $\mathbf{F}_a(t)\equiv \int dS~g(\mathbf{r},t)~
\hat{\mathbf{n}}(\mathbf{r})$, the integration being over the entire bead
surface.  In response to force, the bead moves at an instantaneous
velocity $\mathbf{v} (t)$.  In a viscous medium the components of the
force and velocity are linearly related by
$v_\alpha=\mu_{\alpha\beta}F_\beta$ with $\alpha,\beta = 1,2,3$, where the
mobility tensor $\mu_{\alpha\beta}$ depends on the shape of the
object. For functionalized beads propelled by a comet tail, drag from
tethered filaments in fact far exceeds viscous
drag~\cite{pantaloni:jcb03}.  Whether this would be true for a
stationary bead in an actin cloud is unknown.  However, we are
relatively unconcerned about the exact nature of the drag force: For
small enough velocities, it should be possible to make a linear
expansion in force, with higher-order corrections appearing at higher
speeds.

To complete the description of the dynamics, we need to specify the
time evolution of $g$.  We assume that the rate of change of
$g(\mathbf{r},t)$ is a {\em local} function of $g$, and that it depends on
the net velocity $\mathbf{v}$.  The dependence on $\mathbf{v}$ couples the
field $g$ at different positions on the bead and also provides the
possibility of positive feedback
(e.g. when polymerization is decreased on the front, and enhanced 
on the rear of a moving bead).
The growth rate is then expanded in
powers of $g$ and $\mathbf{v}$, resulting in
\begin{eqnarray}
\label{eqngdot}
\frac{\partial g(\mathbf{r},t)}{\partial t}= \Phi\left[g(\mathbf{r}),\mathbf{v}\right]=-g + a\mathbf{v}\cdot\hat{\mathbf{n}} 
-g^2 +b g \mathbf{v}\cdot \hat{\mathbf{n}} \nonumber \\ 
+d\mathbf{v}\cdot\mathbf{v}+e\left(\mathbf{v}\cdot \hat{\mathbf{n}}\right)^2 - c g^3+\cdots+ \eta (\mathbf{r},t),
\end{eqnarray}
where we have included all terms to second order, and the simplest
cubic term. Note that the vector $\mathbf{v}$ contributes through scalar
forms $v_\perp=\mathbf{v}\cdot \hat{\mathbf{n}}$ and $v^2=\mathbf{v}\cdot\mathbf{v}$.
Such an expansion is presumably valid close to a continuous transition
in which both $g$ and $\mathbf{v}$ are small, and needs to be
self-consistently justified. We have rescaled $g$ and $t$ so that the
coefficients of $g$ and $g^2$ are both $-1$. The sign of the linear
term is negative so that a uniform actin distribution (set to $g=0$
without loss of generality) is stable in the absence of any velocity
coupling, while the sign of the $g^2$ term is unimportant since we may
always consider the dynamics of $-g$ rather than $g$. In addition to
rescaling $g$ and $t$, we may also rescale lengths; a convenient
choice is to set the surface area of the bead to unity. Through
appropriate redefinitions, our previous formulas for the force
$\mathbf{F}_a(t)$ and bead velocity $\mathbf{v}$ remain unchanged. For
analysis and numerical simulations of Eq.~(\ref{eqngdot}) we typically
set $d=e=\cdots=0$, and $c>0$ for stability.  The final term in
Eq.~(\ref{eqngdot}) allows for a stochastic noise $\eta(\mathbf{r},t)$.

In the absence of an external force, Eq.~(\ref{eqngdot}), along with
the formula for velocity
\begin{equation}
\label{eqnvdef}
v_\alpha(t) =\mu_{\alpha\beta} F_\beta(t)=\mu_{\alpha\beta}\int dS ~g(\mathbf{r},t)~ n_\beta(\mathbf{r}),
\end{equation}
fully specifies our model.  We first analyze the case of a spherical
bead, and then generalize its lessons to more complicated shapes.  Due
to its symmetry, the velocity and force are parallel for a sphere, and
$\mu_{\alpha\beta}=\mu\delta_{\alpha\beta}$. In the absence of noise
$\eta$, there is a trivial solution of $g(\mathbf{r},t)=\mathbf{v}(t)=0$,
which corresponds to a static, uniform actin distribution and a
motionless bead. This solution is linearly stable for $a<3/\mu$, but
unstable to dipolar fluctuations for $a>3/\mu$. In the latter case,
the unstable actin fluctuations grow and saturate. Thus
Eq.~(\ref{eqngdot}) allows for a symmetry-breaking transition from a
bead at rest ($\mathbf{v}=0$) to a bead in motion ($\mathbf{v}=const$) as the
parameter $a$ is varied past a critical value $a_c=3/\mu$. This
parameter controls the strength of the positive feedback in the
coupling of the bead velocity to the actin field.

Near this transition we expand $g$ in spherical
harmonics with $g(\mathbf{r},t)\rightarrow g(\Omega,t)=\sum_{\ell
m}g_{\ell m}(t)Y_{\ell m}(\Omega)$, where $\Omega$ represents the solid
angle coordinates $\theta$ and $\phi$.  In this basis
Eq.~(\ref{eqngdot}) becomes
\begin{equation}
\label{eqnglmdot}
\frac{dg_{\ell m}}{dt} = \left(-1+\frac{\mu a}{3}\delta_{\ell 1}\right)g_{\ell m} +\text{nonlinear terms},
\end{equation}
where the nonlinear terms couple different $g_{\ell m}$'s with
coefficients of the form $\int d\Omega
Y_{\ell_1m_1}Y_{\ell_2m_2}Y^{*}_{\ell m}$, $\int d\Omega
Y_{\ell_1m_1}Y_{\ell_2m_2}Y_{\ell_3m_3}Y^{*}_{\ell m},~\cdots$. From the
linear terms in Eq.~(\ref{eqnglmdot}) we see that, near the transition
with $\epsilon \equiv \mu (a-a_c)/3$, $|\epsilon| \ll 1$ and for small
actin fluctuations, the dipolar ($\ell =1$) modes evolve on a slow
time scale $t \propto \epsilon^{-1}$, while the other modes evolve on
a fast time scale $t\propto\mathcal{O}(1)$. Due to their instability
(or near instability) we also expect the dipolar modes to be larger
than the others. We therefore treat the fast $\ell \ne 1$ modes as
adiabatically slaved to the slow $\ell=1$ modes and self-consistently
solve for their amplitudes as functions of the
$g_{1m}$'s. Substitution then yields effective equations of motion for
the slow modes, and by a simple transformation, the bead
velocity~\footnote{Equation~(\ref{eqnvGL}) neglects terms of the form
$\epsilon^n\mathbf{v}^3$ with $n=1,2,3,\ldots$ and similarly at order
$\mathbf{v}^5$. Calculation of these terms by adiabatic elimination
requires more care.}, as
\begin{equation}
\label{eqnvGL}
\frac{d\mathbf{v}}{dt}=\epsilon \mathbf{v}+\frac{27}{5 \mu^2}\left[\left(\frac{\mu b}{3}-1\right)\left(\frac{\mu b}{3}
-2\right)-c\right]\mathbf{v}^3+u\mathbf{v}^5+\cdots,
\end{equation}
with $\mathbf{v}^3\equiv v^2\mathbf{v}$, etc. Thus, near the transition the
bead velocity $\mathbf{v}$ obeys a Ginzburg-Landau equation consistent
with spherical symmetry. When the coefficient of the cubic term is
negative, the transition to nonzero $\mathbf{v}$ is continuous with the
steady state velocity $\mathbf{v}^{*}$ going to zero as $\epsilon
\rightarrow 0^{+}$. In the vicinity of this continuous transition in
parameter space, our analysis above is self-consistently justified and
Eq.~(\ref{eqnvGL}) should provide an accurate description of the
dynamics. The fifth-order coefficient $u$, although not explicitly
presented above, is negative in this region, providing overall
stability.

\begin{figure}
\centering
\includegraphics[width=6.cm]{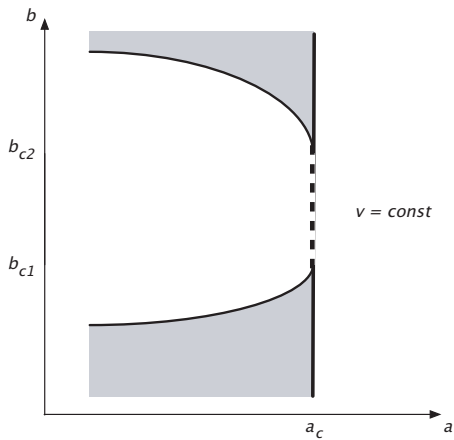}
\caption{\label{figphasediag}Schematic phase diagram for a spherical bead
in the vicinity of continuous transitions for
$a=a_c$ and $b_{c1}<b<b_{c2}$. In the grey region both stationary 
($\mathbf{v}=0$) and moving ($\mathbf{v}=const$) states are locally stable.}
\end{figure}

The phase diagram and critical behavior associated with the
steady-state solutions $\mathbf{v}^{*}$ of Eq.~(\ref{eqnvGL}) are familiar
from mean-field theory in equilibrium statistical mechanics.  The
cubic coefficient is negative when $b_{c1} < b < b_{c2}$, where
$b_{c1}$ and $b_{c2}$ are the two roots of $(\mu b/3-1)(\mu b/3-2)=c$
(recall that $c>0$). In this regime, the transition is continuous with
$\mathbf{v}^{*} \propto \epsilon^{1/2}$. When the cubic coefficient is
positive, the transition is discontinuous.  Tricritical points, at
which $\mathbf{v}^{*} \propto \epsilon^{1/4}$, occur when $b=b_{c1}$ or
$b=b_{c2}$ precisely. Along the line $a=a_c$ or $\epsilon=0$,
$\mathbf{v}^{*} \propto |b-b_{ci}|^{1/2}$ for $i=1,2$ as $b$ varies
outside of the interval $[b_{c1},b_{c2}]$. Finally, in the vicinity of
the first-order transition with $b>b_{c2}$ or $b<b_{c1}$, the
$\mathbf{v}=0$ and $\mathbf{v}=const$ states coexist, both being locally
stable. In the $a$-direction, this region extends over the range
$a^*<a<a_c$ where the ``limit of metastability'' $a^*$ has the $b$
dependence $a_c-a^* \propto (b-b_c)^2$ for $b<b_{c1}$ or $b>b_{c2}$.
The schematic phase diagram for the motion of the bead is shown in
Fig.~\ref{figphasediag}. 

\begin{figure}
\centering
\includegraphics[width=8cm]{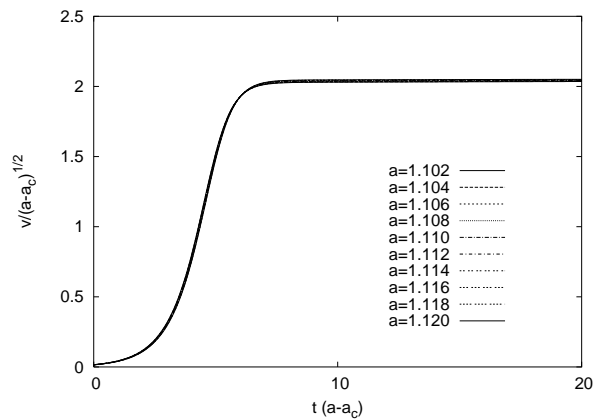}
\caption{\label{figdynamicscaling}Rescaled bead speed versus
rescaled time for $\mu=3$,  $c=0.2$, $b=1.5$ ($b_{c1}\approx 0.83$ and
$b_{c2}\approx 2.17$), and various values of  $a>a_c=1$. The bead begins
(very nearly) at rest at $t=0$ with a small random actin distribution. The
curves are nearly perfectly superimposed.}
\end{figure}

To confirm the analytical results, we numerically simulated
Eq.~(\ref{eqngdot}) with a fourth-order Runge-Kutta scheme.  For
integrations over the surface of the spherical bead, we adapted
routines from the NAG library~\cite{NAG}. We confirmed the phase
diagram in Fig.~\ref{figphasediag}, as well as the various critical
exponents. As an example of the latter, Fig.~\ref{figdynamicscaling}
illustrates the time evolution of velocity for $a>a_c=1$. Starting
from an initial (unstable) stationary state, any small fluctuation
takes the sphere into a moving phase. The characteristic time scale
for this change of state diverges as $\epsilon^{-1}\propto 1/
(a-a_c)$, while the saturation velocity scales as
$\epsilon^{1/2}\propto \sqrt{a-a_c}$.  As indicated in
Fig.~\ref{figdynamicscaling}, the velocity evolution curves obtained
for a number of different $a$ can be collapsed together using the
above scalings of velocity and time.

Let us now consider a rigid bead of arbitrary shape. The acceleration
of the bead can be obtained by taking a time derivative of
Eq.~(\ref{eqnvdef}) and substituting from Eq.~(\ref{eqngdot}). The
analog of Eq.~(\ref{eqnvGL}) for the sphere is then obtained as
\begin{equation}
\label{eqnv}
\frac{dv_\alpha}{dt}=\left(-\delta_{\alpha\beta}+a~\mu_{\alpha\gamma}~\overline{n_\gamma n_\beta}~\right)v_\beta+\cdots+f_\alpha(t).
\end{equation}
At the linear order, properties of the shape are encoded in the tensor
$\overline{n_\alpha n_\beta}\equiv\int dS~ n_\alpha(\mathbf{r})
n_\beta(\mathbf{r})$. The stochastic force $f_\alpha(t)$ partly originates
in the noise in Eq.~(\ref{eqngdot}), and correlations amongst its
components also depend on the shape through the matrices
$\mu_{\alpha\beta}$ and $\overline{n_\alpha n_\beta}$.  The linear
stability of a stationary bead is determined by the
eigenvalues $\{\lambda_i\}$ of the matrix
$\bm{\mu}\cdot\bm{\overline{nn}}$. When $a\lambda_i>1$, the bead is
unstable to motion in the corresponding eigendirection.  The largest
eigenvalue indicates the direction that initially becomes unstable.
Of course, as in Fig.~\ref{figphasediag}, it is possible to have
coexistence of moving and stationary states before the onset of this
instability. In fact, the analysis so far does not rule out coexistence
of several states moving along different
directions~\footnote{Throughout the paper we assume a uniform
coating of the bead. This assures the absence of a constant term in
Eq.~(\ref{eqnv}) since $\int dS n_\alpha(\mathbf{r})=0$ for any closed
surface.}.

It is interesting to note that eigenvalues of $\bm{\overline{nn}}$ are
larger along directions where the shape displays a bigger
cross-section. Thus, ignoring variations in mobility, a pancake shape
prefers to move along its axis (breaking a two-fold symmetry), while a
cigar shape moves perpendicular to the axis (breaking a degeneracy in
angle).  This is opposite to what happens for passive ellipsoids in a
fluid where hydrodynamic effects favor motion along the slender axis.

To see what happens when the stationary bead is linearly unstable, we
need to examine the higher-order terms in Eq.~(\ref{eqnv})~\footnote{To
consider Eq.~(\ref{eqnv}) beyond linear order, we adiabatically
eliminate all modes of $g(\mathbf{r},t)$ except the three-dimensional
subspace corresponding to $\{v_\alpha\}$. This is only reasonable near
the transition when $|a\lambda_1-1|\ll 1$, where $\lambda_1$ is the
largest eigenvalue of $\bm{\mu}\cdot\bm{\overline{nn}}$. The
elimination of fast modes is only partial, in that the components of
$\mathbf{v}$ in directions perpendicular to the unstable eigenvector are
fast, but still retained.}. {\em If the transition is continuous}, its singularities are
determined by the next higher-order term in the expansion.  For the
sphere, symmetry considerations rule out a quadratic  term; the
cubic term leads to the singularities discussed earlier.  However,
quadratic terms $v_\beta v_\gamma$ may appear in Eq.~(\ref{eqnv}) with
coefficients $C_{\alpha\beta\gamma}$ that are related to the shape by
$\overline{n_\alpha n_\beta n_\gamma}\equiv\int dS ~n_\alpha(\mathbf{r})
n_\beta(\mathbf{r})n_\gamma(\mathbf{r})$. If such terms are present, the
velocity will vanish on approaching the transition as $\epsilon$,
rather than $\sqrt{\epsilon}$. To distinguish the two universality
classes along a particular direction, we merely need to ask if the
opposite direction is equivalent. Thus the velocity of a cigar-shape will
vanish as $\sqrt{\epsilon}$, while that of an arrowhead goes to
zero linearly. In the language of dynamical systems, the transition
at $\epsilon=0$ is a pitchfork bifurcation in the former case and a
transcritical bifurcation in the latter~\cite{strogatz:book94}.

Let us now consider the effect of the noise $f_\alpha(t)$ in
Eq.~(\ref{eqnv}).  At a coarse level this resembles a Langevin
equation for the velocity $\mathbf{v}$.  However, since this dynamic
equation is not subject to fluctuation--dissipation constraints, the
resulting probability distributions need not be those familiar from
equilibrium statistical physics.  For example, let us consider an
ellipsoidal shape in the regime where stationary motion is stable, and
higher-order terms in Eq.~(\ref{eqnv}) can be ignored. For a passive
(Brownian) ellipsoid, the velocity fluctuations follow a Maxwell
probability distribution. In this case the center of mass velocity is
equally likely to point in any direction, irrespective of the
orientation of the ellipsoid.  For active fluctuations (e.g. due to
actin polymerization) of the same bead, the probability distribution
of velocities can be extracted from Eq.~(\ref{eqnv}).  Linear
truncation leads to a Gaussian probability distribution,
$P(\mathbf{v})\propto\exp\left(-v_\alpha M_{\alpha\beta}v_\beta\right)$,
but there is no a priori reason for the matrix $M_{\alpha\beta}$ to be
proportional to $\delta_{\alpha\beta}$. Thus generically the velocity
fluctuations of the active bead are non-Maxwellian, and correlated
with the shape of the bead.

In summary, motivated by experiments on actin-based motility, we have
 characterized some of the phenomena that distinguish such
an active system from a passive Brownian particle.  We start with a
highly simplified microscopic model whose central ingredients are a force that is
{\em (a)} locally normal to the surface of a rigid bead, and {\em
(b)} influenced by the velocity of the bead in the surrounding
medium.  These assumptions lead to a macroscopic equation for the
bead velocity  whose coefficients reflect the symmetries in its shape.  
(The reliance on symmetry should increase the validity of the form of this 
equation beyond our particular microscopic model.) 
We note that Eq.~(\ref{eqngdot}) for the dynamics of
the actin distribution is similar to equations that model pattern
formation~\cite{cross:rmp93} or
``self-organization''~\cite{haken:book83} in various contexts. As in
our case, analysis of these systems near threshold can often be
performed by eliminating fast modes.

With the macroscopic equation for the bead velocity we can analyze the
singularities that accompany a continuous transition between stationary
and moving states. Depending on symmetries along the direction of movement, 
the velocity vanishes with a square--root or linear singularity.  We also
examined the fluctuations in velocity of an active bead.  We find
that, generically, the probability distribution is not Maxwellian in
form, and is dependent upon the orientation of the bead.  It would be
interesting to experimentally determine this distribution though
single-molecule experiments on active beads.  In this work we focused
only on the translational motion of a freely moving bead.  Rotations
of a bead, as well as its motion in a trap, are likely to bring in
additional degrees of freedom (position and angular coordinates) that
are coupled to the velocity of the bead. In future work, we would like
to incorporate such effects to form a more complete picture of the
repertoire of motions available to active beads.

\begin{acknowledgments}
The authors thank Arpita Upadhyaya for helpful discussions, and
acknowledge support from the NSF grant DMR-01-18213 (MK), a University
Postdoctoral Fellowship from The Ohio State University (HYL), and an NSF
Graduate Fellowship (AL).
\end{acknowledgments}

\bibliography{draft}

\end{document}